\documentclass[11pt]{article}
\usepackage{amsmath}
\usepackage{amsfonts}
\usepackage{amssymb}
\usepackage{epsfig}
\usepackage{times}
\begin{document}


May 2006

\vskip 1cm
\begin{Large}
\centerline{\bf Inclusive hadronic distributions at small $\boldsymbol{x}$ 
inside jets} 
\end{Large}

\vskip .3 cm

\centerline{Redamy Perez-Ramos
\footnote{E-mail: perez@lpthe.jussieu.fr}
}

\baselineskip=15pt

\smallskip
\centerline{\em Laboratoire de Physique Th\'eorique et Hautes Energies
\footnote{LPTHE, tour 24-25, 5\raise 3pt \hbox{\tiny \`eme} \'etage,
Universit\'e P. et M. Curie, BP 126, 4 place Jussieu,
F-75252 Paris Cedex 05 (France)}}
\centerline{\em Unit\'e Mixte de Recherche UMR 7589}
\centerline{\em Universit\'e Pierre et Marie Curie-Paris6; CNRS;
Universit\'e Denis Diderot-Paris7}

\vskip 0.5cm

{\bf Abstract}: After giving their general expressions valid at all $x$,
double differential 1-particle inclusive distribution
inside a quark and a gluon jet produced in a hard process, together with the
inclusive $k_\perp$ distribution, are given at small $x$ in the
Modified Leading Logarithmic Approximation (MLLA), as functions of the
transverse momentum $k_\perp$ of the outgoing hadron.

\vspace {1mm}
\noindent
{\large\bf Introduction}

This work concerns the production of two hadrons inside  a high energy jet
(quark or gluon); they hadronize out of two partons 
at the end of a cascading process that we calculate in pQCD\cite{PerezMachet}.
Considering this transition as a ``soft'' process is
the essence of the ``Local Parton Hadron Duality'' (LPHD) hypothesis\cite{DKTM}.
We give indeed, in the MLLA scheme of resummation,
the double differential inclusive 1-particle distribution
and the inclusive $k_\perp$ distribution as functions of  the transverse
momentum of the emitted hadrons in the limit that proved successful
when describing energy spectra of particles in jets, that is
$Q_0\approx\Lambda_{QCD}$,
the so called ``limiting spectrum''\cite{DKTM}.
%

\vspace {1mm}
\noindent
{\large\bf The process under consideration}

In a hard collision ($pp$, $p\bar p$ collisions, $e^+e^-$ annihilation)
we consider a jet of half opening angle $\Theta_0$ initiated by a parton $A_0$,
which can be a quark or a gluon, see Fig.1. $A_0$, by a succession of partonic 
emissions  (quarks, gluons), produces a jet of half opening angle $\Theta_0$,
which, in particular, contains the parton $A$; $A$ splits  into $B$ and $C$, 
which hadronize respectively into the two hadrons $h_1$  and $h_2$.
$\Theta$ is the angle between $B$ and $C$.
Because the virtualities of $B$ and $C$ are much smaller than that
of $A$, $\Theta$ can be considered to be close to the angle between $h_1$ 
and $h_2$; angular ordering (AO) is also a necessary condition for this 
property to hold. $A_0$ carries the energy $E$ and gives rise to the 
(virtual) parton $A$, which carries the fraction
$u$ of the energy $E$. $A\to B+C$ occurs with probability $\propto\Phi$, $\Phi$ is the corresponding DGLAP splitting function, $B$ and $C$ carry 
respectively the fractions $uz$ and $u(1-z)$ of $E$; 
$h_1$ carries the fraction $x_1$ of $E$; $h_2$ carries the fraction 
$x_2$ of $E$. One sets $\Theta \leq\Theta_0$.

\vbox{
\begin{center}
\epsfig{file=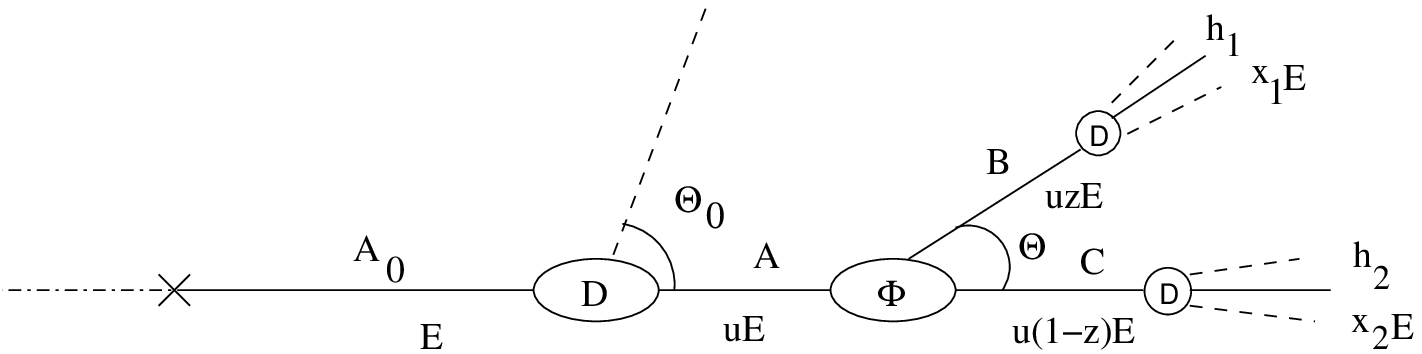, height=2.7truecm,width=9.5truecm}

{\em Fig.~1: process under consideration: two hadrons $h_1$ and $h_2$ 
inside one jet.}
\end{center}
}

Since we are concerned with observables depending on the transverse momentum of 
outgoing hadrons, $k_{\perp}$ should be defined with respect to the jet axis
which is identified with the direction of the energy flow\cite{PerezMachet}.

\vspace {2mm}
\noindent
{\large\bf Double differential 1-particle inclusive distribution at all values of $\boldsymbol{x_1}$}

The energy conservation sum rule, considered together with DGLAP evolution equations
and AO lead the general expression for the double differential 1-particle 
inclusive distribution

\begin{equation}
\frac{d^2N}{dx_1\,d\ln{\Theta}}=
\frac{d}{d\ln\Theta}\sum_{A}\int_{x_1}^1
du
 D_{A_0}^A\left(u,E\Theta_0,uE\Theta\right)D_A^{h_1}\left(\frac{x_1}
 {u},uE\Theta,Q_0\right)
\label{eq:DD}
\end{equation}
which is valid at all $x_1$. $\frac{d^2N}{dx_1\,d\ln{\Theta}}$ in (\ref{eq:DD}) 
is described indeed by the convolution of two fragmentation functions, $D_{A_0}^A$ 
and $D_A^{h_1}$, over the energy fraction
$u$ of $A$. Namely, $D_{A_0}^A$ is purely partonic and describes the probability 
to emit parton $A$ with energy $uE$ off parton $A_0$, moreover, it takes into 
account the evolution of the jet between $\Theta_0$ and $\Theta$; $D_A^{h_1}$ 
describes the probability to produce hadron $h_1$
of energy fraction $x_1/u$ at angle $\Theta$ with respect to the direction 
of the energy flow
inside the subjet $A$. On the other hand, the transverse momentum of $h_1$ 
should be bigger than the collinear cut-off parameter 
$Q_0$ ($uE\Theta\geq Q_0$). Consequently, a lower bound for $\Theta$
is obtained: $\Theta\geq\Theta_{min}\approx Q_0/x_1E$.
%

\vspace {2mm}
\noindent
{\large\bf Perturbative expansion parameter and kinematics}

We conveniently define the variables $\ell_1\!=\!\ln\frac1{x_1}$, $y_1\!=\!\ln{\frac{x_1E\Theta}{Q_0}}$.
In what follows, we use the 
anomalous dimension $\gamma_0$ defined through the running coupling 
constant $\alpha_s$ as
\begin{equation}\label{eq:gammadef}
\gamma_0^2(k_{\perp}^2)\!=\!\frac{2N_c\alpha_s(k_{\perp}^2)}{\pi}\!=\!
\frac2{\beta\ln\displaystyle{\frac{k_{\perp}^2}{\Lambda_{QCD}^2}}}\!\equiv\!
\gamma_0^2(\ell_1+y_1)\!=\!\frac1{\beta(\ell_1\!+\!y_1\!+\!\lambda)}\!=
\!\frac1{\beta(Y_{\Theta}\!+\!\lambda)},\quad\lambda\!=
\!\ln\frac{Q_0}{\Lambda_{QCD}}.
\end{equation}
It determines the rate of multiplicity growth with energy.  
$N_c\!=\!3$ for $SU(3)$, $\beta\!=\!\frac1{4N_c}(\frac{11}3N_c\!-\!\frac43T_R)
\!\!\stackrel{n_f=3}{=}\!\!0.75$ with $T_R\!=\!\frac12n_f$ where $n_f\!=\!3$ 
is the number of light quarks, $\Lambda_{QCD}$ is the QCD scale, $Y_{\Theta}\!=\!\ell_1+y_1=\ln\frac{E\Theta}{Q_0}$
and $Y_{\Theta_0}\!=\!\ln\frac{E\Theta_0}{Q_0}$.
For instance, 
in the LHC environment we can take the typical value $Y_{\Theta_0}\!=\!7.5$ 
$\Rightarrow$ $\gamma_0\!\simeq\!0.4$ by setting $\lambda\!=\!0$ 
($Q_0\!=\!\Lambda_{QCD}$),  $\Theta\!=\!\Theta_0$ and $Q_0\!=\!250\; \text{MeV}$ in (\ref{eq:gammadef}). $\gamma_0$ can be therefore treated 
as the small parameter of the perturbative expansion at MLLA.

\vspace {2mm}
\noindent
{\large\bf Double differential 1-particle inclusive distribution at small 
$\boldsymbol{x_1}$, $\boldsymbol{x_1\ll1}$}

The convolution integral (\ref{eq:DD}) is dominated by $u\approx1$. In order to
obtain an analytical expression of (\ref{eq:DD}), since $x_1/u\ll1$, 
we perform a perturbative expansion in $\gamma_0$
such that (\ref{eq:DD}) gets factorized and derived in the region 
of soft multi-particle production\cite{PerezMachet}
\begin{eqnarray}\label{eq:DDbis}
&& \frac{d^2N}{d\ell_1\,d\ln k_\perp} = \frac{<C>_{q,g}}{N_c}
\frac{d}{dy_1}\tilde D_g(\ell_1,y_1) 
+ \frac{1}{N_c} \tilde D_g(\ell_1,y_1) \frac{d}{dy_1}<C>_{q,g},\\
&& \frac{d}{dy_1}\tilde D_g(\ell_1,y_1) = {\cal O}(\gamma_0)
= {\cal O}(\sqrt{\alpha_s}),\quad
\frac{d}{dy_1}<C>_{q,g} = {\cal O}(\gamma_0^2) = {\cal
O}(\alpha_s).
\label{eq:correc}
\end{eqnarray}

The first term in (\ref{eq:DDbis}) 
is the main contribution to the double differential 1-particle 
inclusive distribution while the second one constitutes its MLLA 
correction of relative order ${\cal O}(\gamma_0)$ (\ref{eq:correc}). $D_A^{h_1}$
in (\ref{eq:DD}) has been replaced by the distribution 
$D_g(\ell_1,y_1)$ in (\ref{eq:DDbis}) 
that describes the MLLA ``hump-backed plateau'' in the 
limit where $Q_0$ can be taken down to $\Lambda_{QCD}$ 
(``limiting spectrum'')\cite{DKTM}.
$<C>_{q,g}$ is the total colour current that describes 
the evolution of the jet between $\Theta_0$ and $\Theta$.
It decomposes into its leading term $<C>_{q,g}^0$ and the 
MLLA correction $\delta<C>_{q,g}={\cal O}(\gamma_0)$\cite{PerezMachet}. 
In Fig.2 we represent $<C>_{q,g}^0$ (straight line) and the full expression 
$<C>_{q,g}=<C>_{q,g}^0\!+\!\delta<C>_{q,g}$ (curve) for two values of $\ell_1$;
$\ell_1=2.5(\text{left}),\;3.5(\text{right})$ in function of $y_1$. 
Two types of MLLA corrections are displayed in this figure. 
Namely, $\delta<C>_{q,g}<0$ is given by the vertical difference between the 
straight and curved lines. The other correction is given by the slope 
of the curve, that is $\frac{d}{dy_1}<C>_{q,g}$ in (\ref{eq:correc}), this one is
large and positive for $y_1\geq1.5$.
For $\ell_1=2.5$ and $y_1\approx1.5$, $\delta<C>_{q,g}$ represents $50\%$ 
of $<C>_{q,g}$ while for
$\ell_1=3.5>2.5$ it gets under control. We are thus allowed 
to set the range of applicability of our soft approximation to
$\ell_1\geq\ell_{1,min}\approx2.5$ ($x_1\lesssim0.08$) $\Rightarrow$ 
$y_1\leq y_{1,max}= Y_{\Theta_0}-\ell_{1,min}=5.0$.
\vbox{
\begin{center}
\epsfig{file=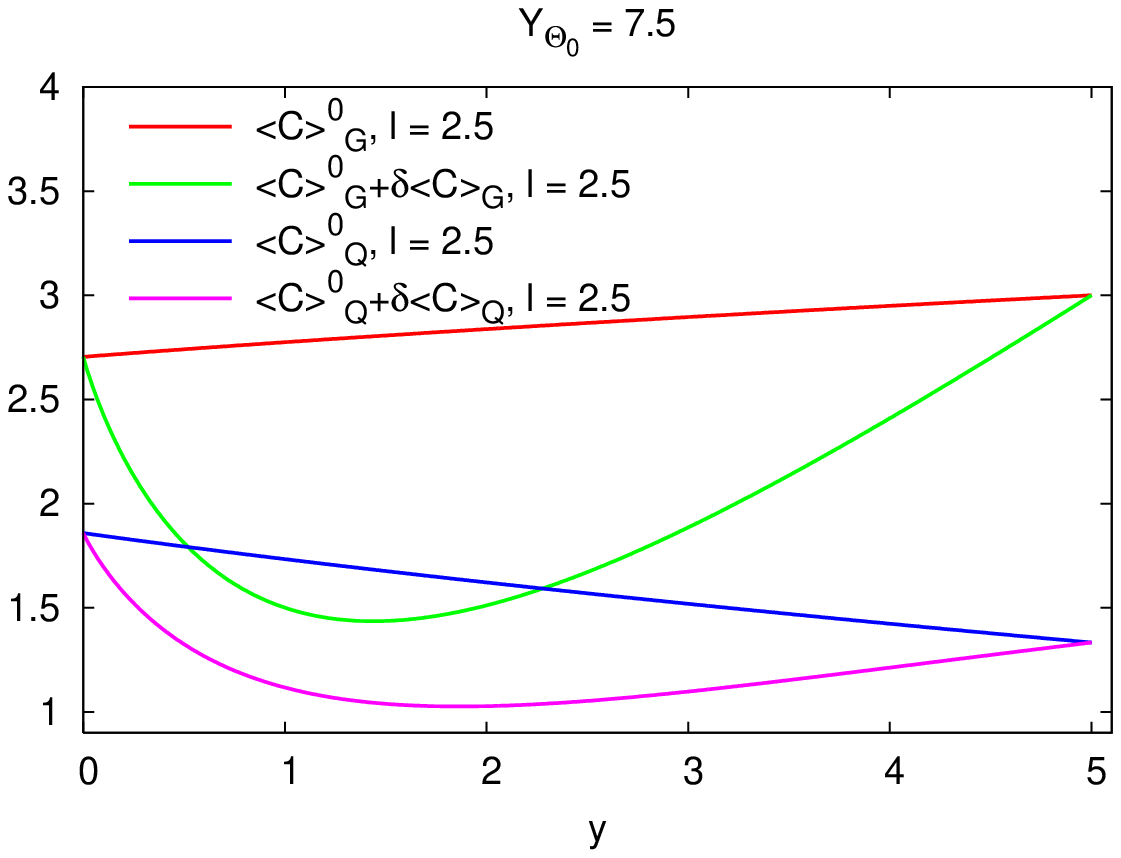, height=4truecm,width=6.5truecm}
\epsfig{file=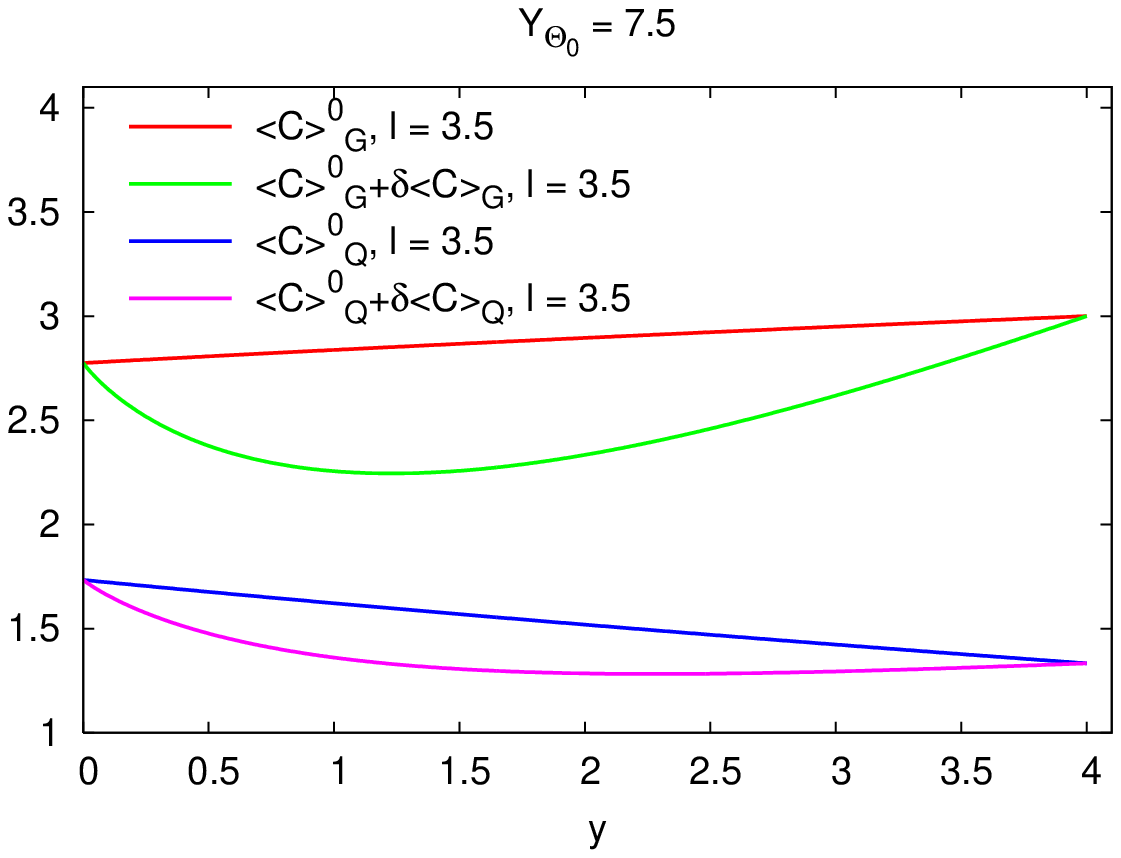, height=4truecm,width=6.5truecm}
\end{center}
\centerline{\em Fig.~2: $<C>_{A_0}^0$ and $<C>_{A_0}^0 + \delta\!<C>_{A_0}$ 
for quark and gluon jets, as functions of $y$,}
\centerline{\em for $Y_{\Theta_0}=7.5$, $\ell=2.5$ on the left and
$\ell=3.5$ on the right.}
}

Furthermore, one should stay in the perturbative regime, which needs 
$y_1\geq1$ ($k_{\perp}>2.72\Lambda_{QCD}\approx0.7\;\text{GeV}$). 
We finally get an estimate
of the range of applicability of the MLLA scheme of 
resummation to be $1.0\!\!\leq y_1\!\leq\!5.0$ in the LHC environment.
In Fig.3 below, we represent the double differential 1-particle 
inclusive distribution (\ref{eq:DDbis}) for $\ell_1=3.5$ in function 
of $y_1$. We compare our results with a naive
DLA-inspired case where one does not take into account the 
evolution of the jet between
$\Theta_0$ and $\Theta$ but sets $<C>_q=C_F$ (quark jet) 
and $<C>_g=N_c$ (gluon jet).

\vbox{
\begin{center}
\epsfig{file=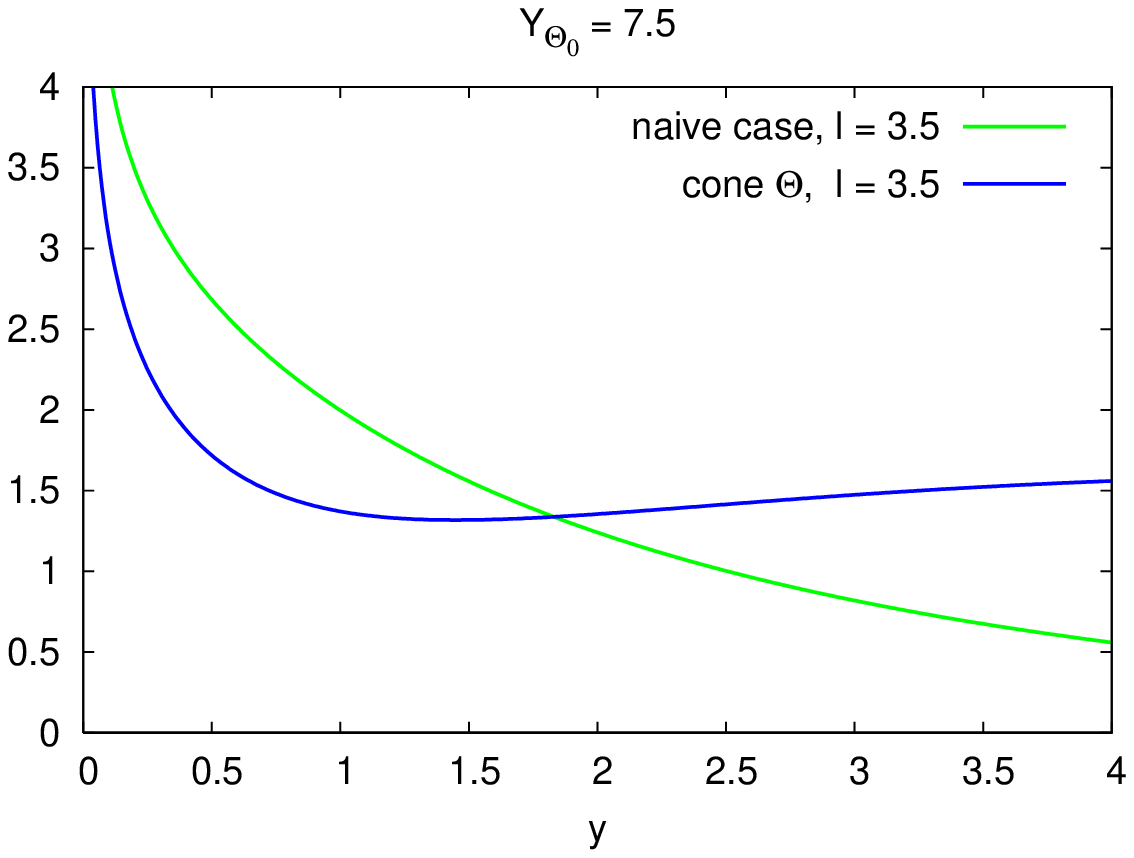, height=4truecm,width=6.5truecm}
\epsfig{file=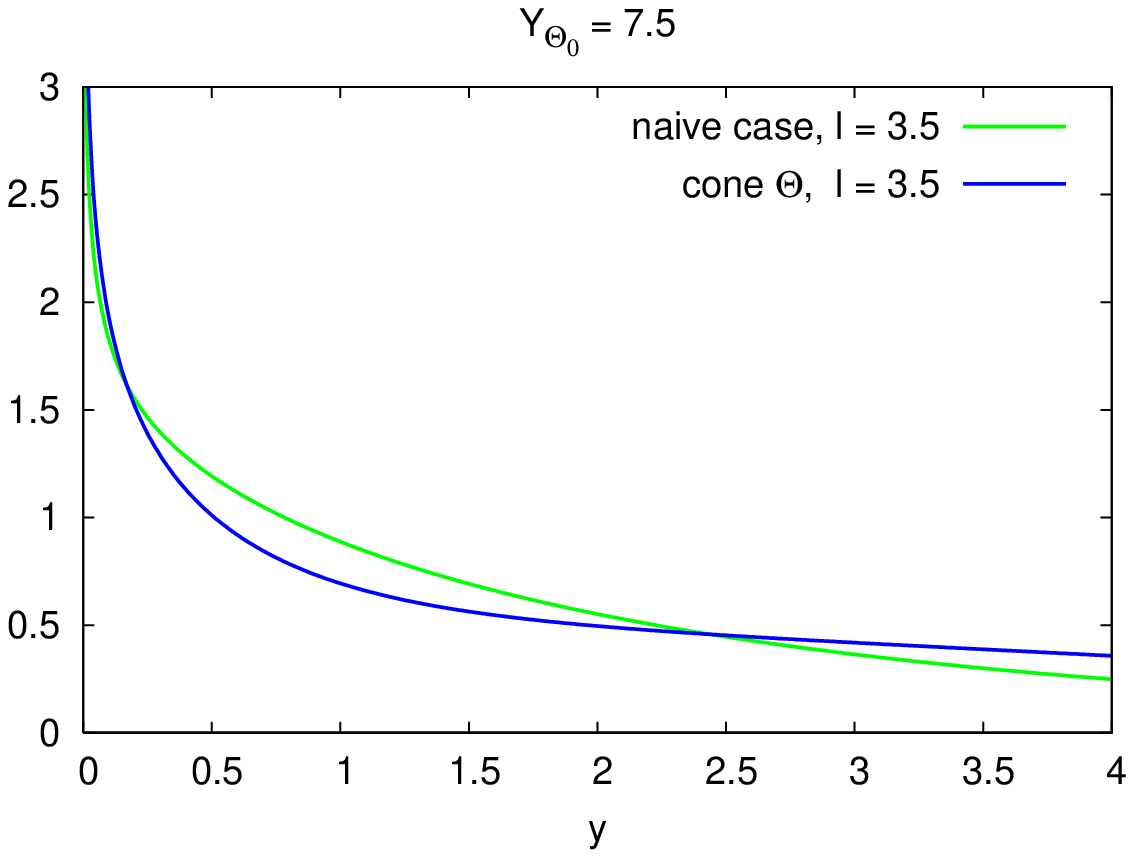, height=4truecm,width=6.5truecm}
\end{center}
\centerline{\em Fig.~3: $\frac{d^2N}{d\ell_1\;d\ln k_\perp}$ 
for a gluon jet (left) and for
a quark jet (right) at fixed $\ell_1=3.5$,}
\centerline{\em  MLLA and naive approach.}
}
For both, the quark and gluon jets, new MLLA corrections arising 
from (\ref{eq:correc}) push up the distribution when $y_1$ increases
as compared with the naive case. On the other hand, it is
enhanced when $y_1\to0$ by the running of 
$\alpha_s(k_{\perp})$ at $k_{\perp}\to\Lambda_{QCD}$.

\vspace {2mm}
\noindent
{\large\bf Transverse momentum inclusive 
$\boldsymbol{k_{\perp}}$ distribution}

Integrating (\ref{eq:DDbis}) over the whole phase space  in the logarithmic sense ($\ell_1=\ln(1/x_1)$)
we obtain the inclusive $k_{\perp}$
distribution inside a quark and a gluon jet\cite{PerezMachet}
\begin{equation}
\left(\!\frac{dN}{d\ln{k_{\perp}}}\!\right)_{q,g}=
\int_0^{Y_{\Theta_0}-y_1}\!\!d\ell_1\!\!
\left(\!\frac{d^2N}{d\ell_1\,d\ln k_\perp}\!\right)_{q,g}.
\end{equation}

We give results for $Y_{\Theta_0}=7.5$ in Fig.4 below and compare with the naive case.

\vbox{
\begin{center}
\epsfig{file=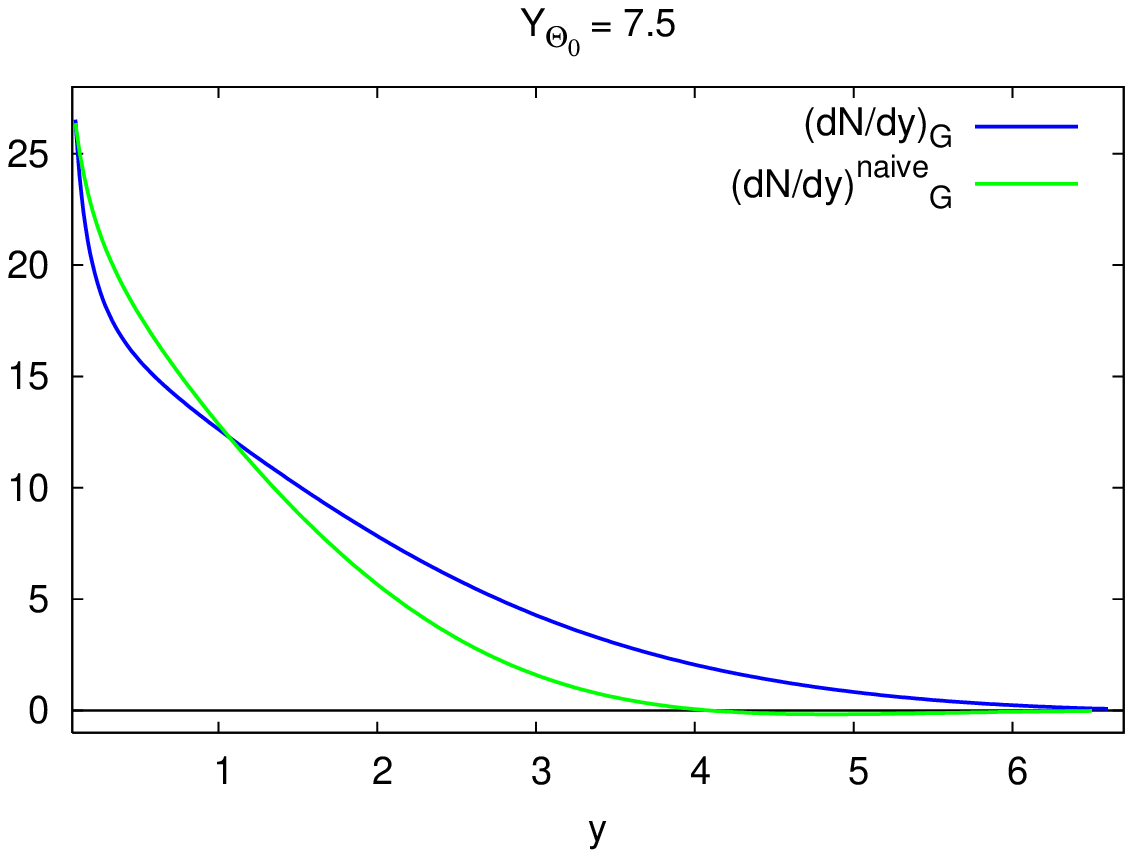, height=4truecm,width=6.5truecm}
\epsfig{file=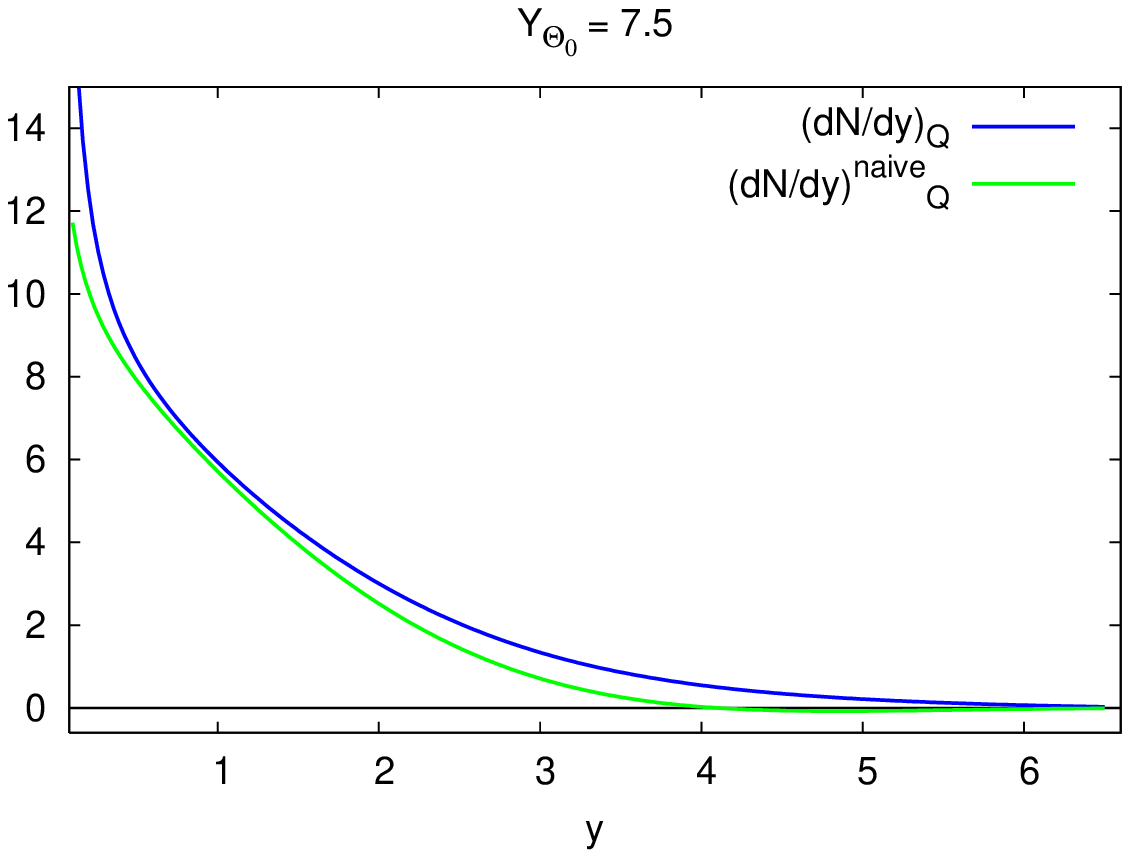, height=4truecm,width=6.5truecm}
\end{center}
\centerline{\em Fig.~4: $\frac{dN}{d\ln k_{\perp}}$  
for a gluon jet,
MLLA and naive approach with enlargement}
}
We observe in particular that the positivity of the distribution 
when $y_1$ increases is restored as compared with the naive case. 
This stems from the global role of MLLA corrections in the range.

\vbox{
\begin{center}
\epsfig{file=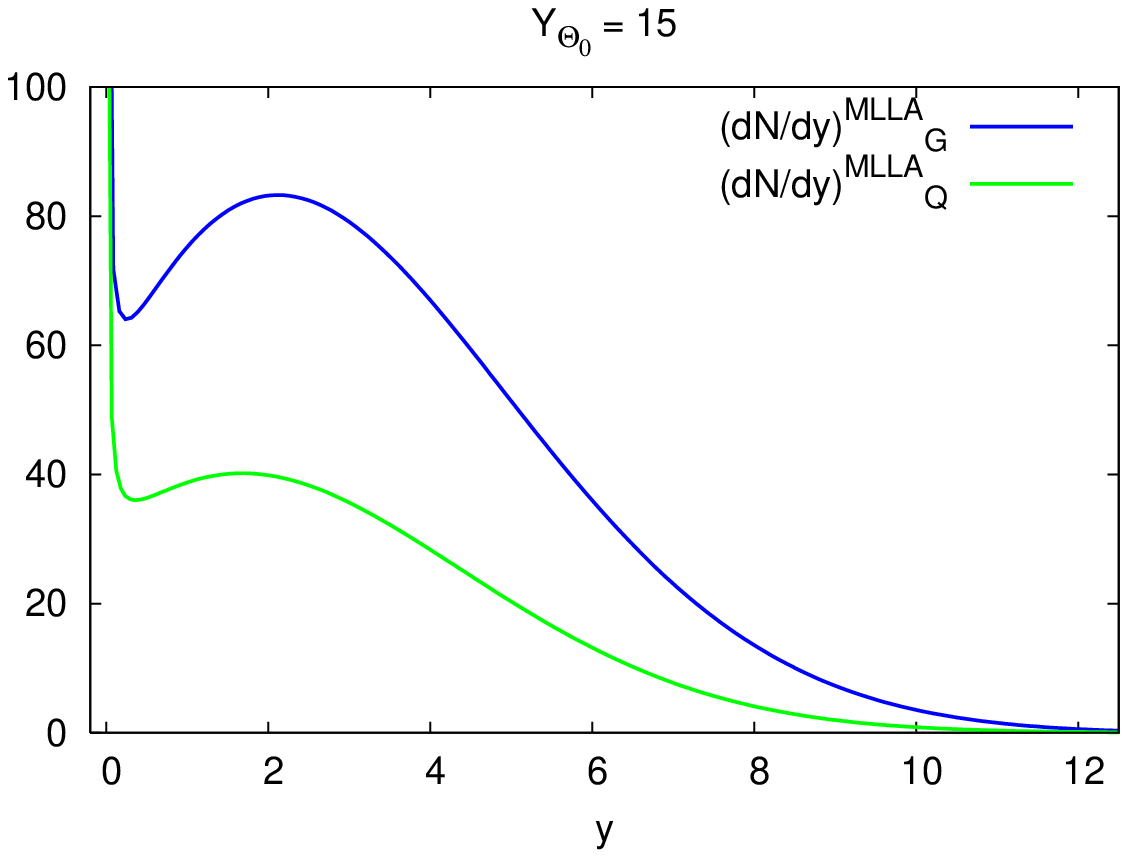, height=4truecm,width=6.5truecm}
\end{center}
\centerline{\em Fig.~5: $\frac{dN}{d\ln k_{\perp}}$  for a gluon jet (blue) and for 
a quark jet (green) at $Y_{\Theta_0}=15.0$
}
}
In Fig.5 above, we represent the evidence of two competing effects at 
the unrealistic value $Y_{\Theta_0}=15.0$. One observes that as $y_1\to0$
the distribution is depleted by QCD coherence effects. Indeed, in 
this region of the phase 
space ($k_{\perp}\to Q_0\approx\Lambda_{QCD}$) gluons are pushed 
at larger angles, they are thus emitted independently from 
the rest of the partonic ensemble. At the opposite, when the
 value $k_{\perp}\approx Q_0\approx\Lambda_{QCD}$ is reached,
the distribution is enhanced by the running of $\alpha_s$.

\vspace {2mm}
\noindent
{\large\bf Conclusion}

Results for the double 1-particle inclusive distribution and the 
inclusive $k_{\perp}$ distribution at small $x$ have been discussed
and displayed. Sizable differences with the naive approach in 
which one forgets the evolution
of the jet between its half opening angle $\Theta_0$ and the 
emission angle $\Theta$ have been
found. The global role of new MLLA corrections is emphasized to recover 
the positivity of the inclusive $k_{\perp}$ distribution. 
On the other hand, at realistic energy scales (LHC, Tevatron, LEP), 
QCD coherence effects are screened by the running of $\alpha_s$ (see Fig.4) which,
furthermore, forbids extending the confidence domain for $y_1\leq1$. 
The range of applicability of the soft approximation has been 
discussed from the analysis of corrections not to exceed 
$y_{max}=Y_{\Theta_0}-\ell_{min}=5.0$;
it is indeed smaller at LEP and Tevatron energies 
(smaller $Y_{\Theta_0}$). Our results will be compared 
with forthcoming data from CDF.


\end{document}